\magnification=\magstep1
\font\bigbfont=cmbx10 scaled\magstep1
\font\bigifont=cmti10 scaled\magstep1
\font\bigrfont=cmr10 scaled\magstep1
\vsize = 23.5 truecm
\hsize = 15.5 truecm
\hoffset = .2truein
\baselineskip = 14 truept
\overfullrule = 0pt
\parskip = 3 truept
\def\frac#1#2{{#1\over#2}}


\topinsert
%
\vskip 1.7 truecm
\endinsert

\centerline{\bigbfont THE LANDAUER FORMULA: A MAGIC MANTRA REVISITED$^*$}
\footnote{}{* Invited talk, XXVI International Workshop on Condensed
Matter Theories, Luso, 2002}

\vskip 20 truept

\vskip 8 truept
\centerline{
{\bigifont Mukunda P. Das}
{\bigrfont and}
{\bigifont Frederick Green}
}
\vskip 8 truept
\centerline{\bigrfont Department of Theoretical Physics}
\vskip 2 truept
\centerline{\bigrfont Research School of Physical Sciences and Engineering}
\vskip 2 truept
\centerline{\bigrfont The Australian National University}
\vskip 2 truept
\centerline{\bigrfont Canberra ACT 0200, Australia}
\vskip 2 truept

\vskip 14 truept

\vskip 1.8 truecm

\centerline{\bf 1. INTRODUCTION}
\vskip 12 truept

\centerline{\it a. Prolog}
\vskip 8 truept

In 1957, Rolf Landauer published a different and -- to some
of the leading transport gurus of the epoch -- iconoclastic
interpretation of metallic resistivity[1].
Landauer envisioned the current, rather than the electromotive
voltage, as the stimulus by which resistance is manifested[2].
The measured voltage is simply the macroscopic
effect of the carriers' inevitable encounters with the localized
scattering centers within a conductor. Around any such scatterer
the carrier flux resembles a {\it diffusive flow}, set up by the
density difference between the upstream and downstream electron
populations. In this purely passive scenario, energy dissipation
does not enter.

The Landauer theory describes electron transport in an environment
of scatterers that are {\it purely elastic}. As such
it is not able to address the dynamical mechanisms of energy
dissipation that characterize transport. This is despite the fact that
Landauer theory, like any other description of conductance,
must satisfy the fluctuation-dissipation (Johnson-Nyquist) theorem
at some level. The theorem implies that dissipation is always
present whenever there is resistance. It is an inescapable element
in every theory of conductance. A properly formulated
theory will describe dissipation explicitly, in physical detail.

This paper reviews the conceptual structure of the Landauer model
of electron transport, in the light of what has long been accepted as the
canonical description of conductance and fluctuations: quantum kinetics.
Through some straightforward instances, we discuss the lack of
a clearly discernible correspondence between Landauer theory
on the one hand, and standard microscopics on the other.

From our microscopic critique it follows that one must consider,
longer and harder than before, how Landauer's bold insight
can be said to express its microscopic integrity. There is no question
that Landauer, with his foresight and determination,
inspired and guided the growth of mesoscopic physics
as very few other personalities could have done. However,
establishing a substantive connection with canonical microscopics
remains a serious task, whether or not that fact is widely
acknowledged.

\vskip 12 truept
\centerline{\it b. The Rise of Landauer Theory}
\vskip 8 truept

Landauer has recounted[3] how his ideas languished until
their resurrection, and bold reinterpretation, by a new generation of
less hide-bound theorists. The main early objection seemed to be
his emphasis on the {\it localized} action of the scattering
impurities in opposing current flow, in an era when the theoretical
dogma held that local effects could never be individually probed;
all that one could (and should) do was to compute a configurational
average over ensembles of samples, with a spread of scattering-center
distributions. A clear account of the configurational
viewpoint is in the text by Doniach and Sondheimer[4].

All this changed radically in the 'eighties, with the
advent of truly mesoscopic sample fabrication. It was now possible to
study not bulk, coarsely grained ensembles, but individual samples with
individual spatial arrangements of scatterers. Moreover, the phase
coherence of the carriers could now be preserved over the much
shorter lengths of the samples.

Overnight, the Landauer model came into its own.
A new dogma was promptly declared (though not primarily
by Landauer himself): transmission is all.
One result of this sea change has been that standard,
thoroughly established and powerful microscopic methods,
such as Boltzmann kinetics
and the quantum mechanical Fermi-liquid and Kubo theories,
now languish in relative disuse within mesoscopics.
Those papers that apply such methods to the subject
tend to be obscured in the noise
(we have reviewed several such works elsewhere[5]).

For one thing, the older approaches are much too
labor-intensive to suit the milieu of instant results. For a second,
it is taken -- on faith alone -- that they can do little
else than corroborate Landauer's far more compact phenomenology.
For yet a third, all mesoscopic transport was now to occur solely
by coherent elastic transmission, to all intents dissipationless
(even when the fluctuation-dissipation theorem clearly states otherwise).

It is fair to say that the eclipse of standard microscopic methods in
mesoscopics has no particular basis in logic. It is fashion-driven.
Fashion dogma being ultimately less hard-wearing than good physics,
the vogue {\it circa} 1957 did not last. Will the current one endure?$^1$
\footnote{}{$^1$ Landauer reprinted his 1957 paper in 
J. Math. Phys. {\bf 37}, 5269 (1996), with a self-explanatory
commentary:  `` ...IBM Journal of Research and Development,
is not all that easily located in 1996. As a result the frequent
citations to it often assign content to that paper
which does not agree with reality.''}

Meanwhile, in other disciplines where short-range coherent phenomena
are equally pivotal (such as nuclear matter), microscopic analysis
remains not only indispensable but goes steadily from strength to strength,
both in its power and sophistication. For an overview of modern
developments in many-body quantum kinetics, see the volume
edited by Bonitz[6]. We now engage our subject technically.

\vskip 20 truept
\centerline{\bf 2. ISSUES IN MESOSCOPIC TRANSPORT}
\vskip 12 truept

\centerline{\it a. Importance of the Landauer Formula
in the Mesoscopic Literature}
\vskip 8 truept

Landauer's suggestion that conduction is at heart transmissive dominates the
realm of mesoscopic physics[1]. His formula, derived for noninteracting
electrons in a one-dimensional (1D) uniform conductor, has wide
application to a variety of physical
systems: quantum wires, quantum-Hall edge states, quantum point
contacts, and carbon nanotubes. There is a current consensus
that the physics and technology of transport in nanodevices are fully
understood by the seminal work of Landauer and its extensions, and
that his model supersedes all others.

We will clarify a number of important points, which
have taken on a mythical significance in the electron-transport theory
of mesoscopic conductors. Some of the arguments that we present
may be at odds with the consensual Landauer formalism in terms
of fundamental physics. In our microscopic formulation, the Landauer
formula occurs as a highly idealized case. There is an important
departure here: as for any standard {\it microscopic} treatment, our
description of the physics of dissipation is explicit. So it must be.
In Landauer-like descriptions, it remains a congenitally murky issue.

Let us begin with a simple derivation of the Landauer conductance,
after Imry and Landauer[7].
A long, uniform 1D wire is connected to two large leads held
at different chemical potentials $\mu_1$ and $\mu_2$, where all
energies are measured from a global zero level. Electrons
flow from the higher potential $\mu_1$ to the lower one $\mu_2$
(that is, the density $n(\mu)$ is assumed to increase monotonically).

A disturbance $\delta n$ of the equilibrium electron density
across the ends of the wire defines the current-density response as

$$
J = (-e) v {\delta n\over 2}
\equiv (-e) {v\over 2}
{\partial n\over \partial E} \delta E
= (-e) {v\over 2}{\left( {2\over \pi} D(E) \right)} (\mu_2-\mu_1),
\eqno(1)
$$

\noindent
where $v, E$ are the Fermi velocity and energy respectively;
$D(E)$ is the 1D density of states.
Note two points:

\item{(i)}
the change in the physical carrier density
$\delta n$ must be reduced by an {\it ad hoc} factor of two.
That is because only ``left-to-right'' moving
equilibrium carriers are understood to convey
current from source to drain (an equal number of
equilibrium electrons at the source flows the ``wrong'' way).
Conversely, only ``right-to-left'' moving ones can
carry the counter-current. Their net sum is $J$.

\item{(ii)}
To expand perturbatively about the Fermi level,
we must be in the extreme degenerate limit
where all relevant energies are much less than $E$.
The argument will not hold for classical ballistic carriers.

Next, by identifying $\mu_1-\mu_2$ with the drop in
electron potential $eV$
across the asymptotic source and drain leads,
and by replacing $D(E)$ with its expression in terms of
$v$, we get

$$
J = (-e) v {D(E)\over \pi} (-eV)  
= e^2 v {\left( {1\over \pi\hbar v} \right)} V
= {\left( {2e^2\over h} \right)} V.
\eqno(2)
$$

\noindent
The Landauer conductance in 1D follows: 

$$
G = \lim_{V \to 0} {\left\{ {J\over V} \right\}}
= {2e^2\over h} = 7.75\times10^{-5} {~}{\rm Siemens}.
\eqno(3)
$$

\noindent
The important quantum ingredient is the one-dimensional density of
states $D(E)$. Its proportionality to $1/v$ cancels
$v$ in the numerator and reproduces the ``universal'' conductance
quantum $2e^2/h$
{\it irrespective of the length or material of the wire}.

Landauer gave a wider interpretation of conductance quantization in terms
of independent eigenchannels, through the transmission matrix.
However, the real question is: What causes {\it dissipation}
in a ballistic (collision-free) wire? That question remains
without a definitive answer, though imaginative explanations have
not been wanting[8-11].

The next serious point concerns the controversial use of several
different chemical potentials within a driven system which is,
{\it ipso facto}, (a) not in thermodynamic equilibrium[3], and
(b) part of a topologically closed electrical circuit.
The common claim that such a mismatch is absolutely required for
{\it any} description of mesoscopic transport, is more than misleading.
It is frankly incorrect, as Kamenev and Kohn have demonstrated
through their completely orthodox quantum-kinetic derivation of Eq. (3)[12].
The thermodynamically erroneous idea of transport-induced
mismatch in chemical potentials continues to propagate in the literature.
It is widely employed in molecular-transistor studies; see for example
Ref. [13].

\vskip 12 truept
\centerline{\it b. Contact Effects in Quantized Conductance}
\vskip 8 truept

Imry[8] has given an intuitive understanding of how the finite
resistance of a collision-free ballistic wire arises; it is due
to the bottleneck between the rich densities of states in leads and
interfaces, and the sparse one for the few open channels
that are available in the wire itself.
He emphasizes the central role of the reservoirs in the process
of dissipation. Since the power dissipated in conduction
is $P = IV = GV^2$,
it is certainly accessible in a sample whose resistance
$G^{-1}$ is of order 13 k$\Omega$ and no-one doubts its reality.

If one tries to follow logically the physics of the
classic derivation of $G$ reproduced in Eq. (3), 
it is far from clear how dissipation {\it in the reservoirs} mediates
the finite resistance for the ballistic channel, unless there is some
manifest role for the reservoirs within the proof of the formula.
There is none, of course; it is a case of ``acting at a distance'',
with the actors not even allowed within sight of the stage.

It is surprising, to say the least, that the result for $G$
does not depend in any way on material parameters 
or the structure of the wire-lead contacts. After all, the Imry
picture still asserts that the role of remote dissipation in the leads
(via inelastic collisions) is central to ballistic transport.
The fundamental incompleteness of this approach becomes evident.

There is universal agreement that dissipative effects are
essential to stabilize transport, ballistic or not.
In subsequent accounts of Landauer theory, they are invoked
as an essential function of the sample's contacts with the
macroscopic leads. Yet the same theory makes no room for dissipative
processes within the core mathematical derivation
of conductance. In our view this is an unsatisfactory
state of affairs for a serious mesoscopic transport model.

\vskip 12 truept
\centerline{\it c. Experiments}
\vskip 8 truept

Recently, in groundbreaking work, de Picciotto {\it et al}.[14] 
demonstrated electronic transport in a nearly ideal, ballistic
quantum wire. The wire's remarkably fine construction rendered
electron flow intrinsically resistanceless.
In this device any resistance appearing in two-terminal measurements$^2$
\footnote{}
{$^2$ Two-terminal measurements probe both
voltage and current across the macroscopic leads that feed
the sample. Four-terminal measurements are those
in which the voltage is measured at distinct
probes on, or immediately adjacent to, the wire itself.
See below.}
is ascribed to the current
contacts, or interfaces, of the wire with its reservoir leads.
Irrespective of the perfectly {\it resistanceless}
quantum wire connected to
the ideal leads, the contact resistance per open channel
is very near the rather substantial value of
$(h/2e^2) \approx 12.9$ k$\Omega$.

The de Picciotto {\it et al}. study[14] is a tour de force.
It goes to unprecedented lengths to fabricate perfectly
ballistic 1D structures,
minimally influenced by extraneous effects.
Even so, the surprise is that their raw results for conductance
quantization clearly show systematic deviations from the perfect
Landauer values. This cannot be so, according to theory, unless
the transmission in the sample is {\it less} than
perfectly ballistic and/or dissipationless.

The above experiment, unquestionably the
cleanest yet done, can be directly compared with
the much earlier data from a pioneering
study of 1D ballistic conductance: van Wees {\it et al}.[15].
This was the first to report perfect quantized conductance
in integer units $n(2e^2/h)$. A careful reader of Ref. [15]
will note that, from the measured resistance, a lead resistance
of 400 $\Omega$ was subtracted (an error of 3\% in the value of $h/2e^2$).
Once plotted, the thus-corrected conductance exhibited
ideal quantized steps, right up to $n=12$.

Our central question is: What causes dissipation in the
ballistic channel of van Wees {\it et al}.?
If, as argued by Imry[8], the perfectly quantized resistance
truly represents the {\it whole} of the ballistic wire's contact
``bottleneck'' to its entire electrical environment,
what is the physical significance of subtracting 400 $\Omega$?
If it were somehow part of the contact resistance
to the sample, its removal would negate a contribution that
should be integral to the perfectly quantized value.
If not, then it must be an experimentally distinguishable
effect lying beyond the all-inclusive Landauer-Imry explanation of
the contact resistance. No circuit-specific origin for
the 400 $\Omega$ was identified by van Wees {\it et al}.[15].

The point is that the contact resistance is not an additive
lumped-circuit parameter (albeit it dissipates electrical
power just as well). It cannot be analyzed as such.
A mesoscopic device does not possess the well-defined boundaries
to permit the additivity that we know from the world of
discrete macroscopic components.

\eject
\centerline{\bf 3. FURTHER ANALYSIS}
\vskip 12 truept
 
\centerline{\it a. Two-Terminal versus Four-Terminal Measurements}
\vskip 8 truept

After several confusing debates in the literature, see for example [8],
the current consensus is that the meaning of conductance
depends on the type of measurement that is
done. Points {\it a-c} of the previous Section relate
to two-terminal measurements, where
the current leads are also the voltage leads. The measured resistance of
a two-terminal device cannot reveal the intrinsic resistance of the
wire (that is, its ``true'' value considered as a discrete component,
presumably in series with the contact resistance).

To gain insight into the intrinsic resistance,
one has to make a four-terminal measurement by putting two
extra voltage leads (probes) to the wire located between the
reservoir contacts. These voltage probes must be noninvasive; they
cannot interfere with the passage of current in the wire.
The voltage drop between these probes, divided by the current
through the wire, should give the resistance of the wire.

In the Landauer-B\"uttiker formulation [8,9,10] this resistance
is given by $G^{-1}$, where now

$$
G = {2e^2\over h} { {\cal T}\over {1-{\cal T}} }.
\eqno(4)
$$

\noindent
The transmission factor ${\cal T}$ gives the transparency of
the barrier represented by the wire.
In this picture ballistic transmission, or ideal reflectionless
elastic ``scattering'' (${\cal T} = 1$), produces ideal (infinite)
conductance, as envisaged long ago by Bloch and others.

Realization of the foregoing measurement scenario -- the attainment of
${\cal T} = 1$ -- was the express goal of de Picciotto {\it et al}.[14].
They reported the vanishing of the intrinsic four-probe
resistance, over a wide range of gate voltage. As we have already noted
in Sec. 2, the experiments of de Picciotto {\it et al}.,
show a two-terminal resistance that is nonuniversal; that is,
${\cal T} < 1$, while the four-terminal resistance is ideal (zero),
implying ${\cal T} = 1$. A primitive understanding of 
Landauer theory might lead to the conclusion that the two-terminal
and four-terminal measurements jointly harbor a contradiction.

\vskip 8 truept
\centerline{\it b. Scattering Theory and Transmission}
\vskip 8 truept

The core of the Landauer theory is the assertion that
{\it transmission is conductance}. Transmission is
to be computed from the
quantum mechanics of elastic single-particle scattering$^3$
\footnote{}
{$^3$ At this simple level of description, any attempt to include
inelasticity (dissipation) destroys the unitarity
of the Schr\"odinger evolution.
The wave function no longer conserves probability.}.
A problem arises when the wire size is of the same order as
the inelastic scattering length. In that case the transmission
properties of the wire are strongly affected by the leads.
The current-carrying state has a dissipative component,
spoiling its coherence. Even when the wire is perfectly
ballistic, its contacts with the leads will induce a certain amount of
decoherence. Therefore the zero-resistance state of the four-probe results
(see, for example, Fig. 2 of Ref.[14]) remains unexplained.

In principle, scattering theory is microscopic. In practice,
its use in Landauer-based models is as a prop to
phenomenology, rather than the reverse as in most other models.
The elements of the scattering matrix
are not extracted microscopically, but are left
for some separate calculation to feed into the Landauer theory.
In thus ``outsourcing'' its physics, such a framework
does little to explain the key details of real devices. 

An important point is that the scattering theory invoked by
Landauer, B\"uttiker, and Imry
is a one-particle theory which considers
only elastic scattering and ignores altogether the
inelastic processes vital to the stability of mesoscopic
transport. We recall that dissipation
is the unique outcome of inelastic scattering, and that dissipation
through any resistance is mandatory (the Johnson-Nyquist theorem).
On that score the occurrence of a finite resistance in
a family of models that are {\it purely elastic},
and in which the action of dissipation in the leads
(conceded by all to be essential) is so vestigial
as to be totally invisible in the formalism,
is an astonishing result.

\vskip 20 truept
\centerline{\bf 4. MICROSCOPIC RESOLUTION}
\vskip 12 truept
 
Dissipation is always a many-body problem. It is mediated by electron-phonon
and electron-electron interactions, and many others besides.
These processes unfold side by side with, say, impurity or
barrier scattering which are well described as elastic one-body
effects.

Whether classical or quantum, unless there is an external agent,
a particle continues to travel
unhindered. If there are correlations among the particles, their freedom
and independence are at least partially lost.
This dynamic is readily followed within Kubo's formulation
of linear-response theory:
microscopic correlations in the current response will
produce, directly and transparently, a finite conductivity[16]. 

\vskip 8 truept
\centerline{\it a. Linear Response and the Many-Body Kubo Formula}
\vskip 8 truept

The Kubo theory is a theory of the electrons' full density matrix,
not of the far more restrictive set of one-body wave functions.
We briefly consider the core structure of the Kubo conductance formula.
Complete details, including the explicit microscopic construction
of the formula from both elastic {\it and} inelastic scattering
processes, are in Ref.[16]. The electrical conductivity is given by

$$
\sigma(t) = {ne^2\over m^*}\int^t_0 {\cal C}_{vv}(tí) dtí 
\eqno(5)
$$

\noindent 
The velocity-velocity correlation function is 

$$
C_{vv}(t) = {{\langle v(t)v(0) \rangle}\over
             {\langle v(0)\rangle}^2} 
\sim \exp(-t/\tau_m)
\eqno(6)
$$

\noindent
where the expectations are taken in the equilibrium state (this gives the
leading linear term in the expansion of the nonequilibrium response),
and $\tau_m$ characterizes the dominant decay of the correlations.
This parameter includes, on an equal footing, the microscopic
contribution from every physically relevant collision mechanism[16].
In the long time limit,

$$
\sigma \to {ne^2\tau_m\over m^*}.
\eqno(7)
$$

\noindent
This is the celebrated Drude formula.
In one dimension, the density in terms of the Fermi
wavenumber is $n = 2k_{\rm F}/\pi$.
The conductance over a sample of length $L$ becomes

$$
G \equiv {\sigma\over L} = {2k_{\rm F}e^2\over \pi L m^*} \tau_m
= {2e^2\over h}{\left( {2\hbar k_{\rm F}\over L m^*} \tau_m \right)}
\equiv {2e^2\over h}{\cal T}_{\rm K},
\eqno(8)
$$

\noindent
in which the transmission coefficient ${\cal T}_{\rm K}$
(K for Kubo) is now proportional
to the ratio of the effective scattering length $v_{\rm F}\tau_m$
to length $L$.

All of the dissipative many-body effects have been
embedded within $\tau_m$, as well as any elastic
impurity scattering. The interface physics is
incorporated into the microscopic Kubo conductance
as fully and directly as the physics of the device itself.

There is nothing in Eq. (8)
that precludes conformity with the Landauer formula, Eq. (3).
Unfortunately, the reverse is not true. This is plain from
Eqs. (1) and (2), its usual phenomenological derivation.
Quite unlike the analytical structure of the
Kubo formula, there is no provision for the many-body physics
{\it at the interfaces} that explicitly generates the dissipation.
Kubo, by contrast, includes all of the physical detail that
this problem demands.

\vskip 8 truept
\centerline{\it b. The Landauer Formula without Landauer's Assumptions}
\vskip 8 truept

The microscopically consistent Kubo derivation of
Eq. (3) by Kamenev and Kohn[12] is a landmark.
It lays to rest the myth that response theory can work only
within the thermodynamically unfounded leaky-reservoir paradigm
(where a mismatch of source and drain chemical potentials
is fancied to accompany the current).
However, their derivation was for a closed mesoscopic 1D loop not
subject to the major dissipative effects that govern an open mesoscopic
system. We now show that the Kubo formula also works just as well
for open systems, and that it recovers the ideal Landauer-conductance
expression as a (very) special case of Eq. (8) ``in the open''.

Let us consider a simple model for the behavior of ${\cal T}_{\rm K}$.
The wire is uniform; so are the driving field and carrier distribution
within. At a distance $L$ apart lie the interfaces where the current
is, in effect, injected and extracted. $L$ is not a geometrically
defined quantity. Rather, it characterizes the dynamical processes
for the open mesoscopic system (somewhat abstractly but quite
specifically; see below).
Note that it is the injection and extraction of the current
that explicitly energizes the system. There is no
appeal to chemical-potential differences in any way, shape, or form.

The interfaces are regions of strong
elastic scattering with impurities in the leads
(relaxation time is $\tau_{\rm el}$)
and strong {\it dissipative} interactions with the background modes
excited by the influx and efflux of carriers from the current source
(relaxation time is $\tau_{\rm in}$)$^4$.
\footnote{}{$^4$ From the viewpoint of a carrier inside
the 1D wire, its dynamical evolution in the presence of
the open leads is Markovian, but over a length scale
set by $L$ if the wire is truly collisionless (ballistic).}
The scattering mechanisms are independent, so that the
Matthiessen rule applies:

$$
{1\over \tau_m} = {1\over \tau_{\rm el}} +  {1\over \tau_{\rm in}}.
\eqno(9)
$$ 

\noindent
The mean free path (MFP) associated with the elastic collisions is
obviously $L$, since by hypothesis that is the size of the
impurity-free region. Therefore $\tau_{\rm el} = L/v_{\rm F}$.
By the same token the MFP for inelastic scattering cannot be {\it greater}
than $L$, though it may be less. Hence

$$
\tau_{\rm in} \leq \tau_{\rm el} = L/v_{\rm F}.
\eqno(10)
$$

\noindent
We conclude that

$$
{\cal T}_{\rm K}
= {2v_{\rm F}\over L}{\left( {{\tau_{\rm el} \tau_{\rm in}}\over
            {\tau_{\rm el} + \tau_{\rm in}}} \right)}
= {2\tau_{\rm in}\over {\tau_{\rm in} + L/v_{\rm F}}}
\leq 1.
\eqno(11)
$$

It is the explicit competition between the elastic processes
in the mesoscopic system (whose very fabrication guarantees that
the characteristic length $L$ will be equal to the elastic MFP),
and the dissipative processes (hopefully restricted to the
current injection/extraction areas bounding $L$,
but also liable to intrude into the interior)
that determines the physical, and measurable, transmission
through the sample.
A full-scale Kubo analysis would clarify the physics in all its microscopic
detail. Nevertheless, the essence of it is already in Eqs. (8)-(11).

What is the optimum outcome for Eq. (11), and what
does it yield for the conductance? The maximum value of
${\cal T}_{\rm K}$ is unity, and it is attained precisely when

$$
\tau_{\rm in} = \tau_{\rm el} = L/v_{\rm F}.
\eqno(12)
$$

\noindent
In other words, no inelastic events intrude into the
core of the wire; they all occur at the interfaces.
From Eq. (8) one easily discerns the corresponding value of $G$
for this open, maximally ballistic 1D wire.
It is the Landauer conductance $2e^2/h$.

Which assumptions have been made in common with Landauer?
There are two.

\item{$\bullet$}
That the channels available in 1D transport are
discrete, and sufficiently apart in energy that each
can be treated independently;

\item{$\bullet$}
That the conductor is internally uniform over most
of its operative length.

\noindent
What have we NOT assumed on the way to Landauer's
rightfully celebrated result?

\item{$\bullet$}
{\it That transport is exclusively due to an energy
drop between carrier reservoirs at different (effective) densities};

\item{$\bullet$}
{\it That coherent elastic transmission is the sole mechanism
that should be included in the formula};

\item{$\bullet$}
{\it That dissipation in an open conductor,
though acknowledged to be vital
(simply to save the Johnson-Nyquist theorem), is a remote asymptotic
effect in the reservoirs with no immediate role in conduction}.

\vskip12 truept
\centerline{\it c. Quantum Kinetics and Quantized Conductance}
\vskip 8 truept

Both the Kubo and Landauer formulae, though radically different
in philosophy, presuppose an external driving force that is weak
enough to let one linearize the nonequilibrium transport equation.
In practical mesoscopic devices, there is no guarantee that
typical internal fields are small enough to justify linear
response. As a prosaic example, we cite the case of high-mobility
transistors with a strongly quantum-confined heterojunction channel.
At optimum operation in a microwave-amplifier circuit, such
structures must reliably sustain electric fields of order
50 kV/cm over their active region. The upper bound for linear
response is much lower, at best 3 kV/cm in GaAs.

There is an evident need to extend transport theory to the nonlinear
regime, while retaining all the quantum effects (not
least, many-body correlations) that impart unique properties to
mesoscopic structures, making them desirable for novel applications.
This is done most naturally within a quantum kinetic
approach[17,18].
We briefly review our basic kinetic treatment
of nonequilibrium transport which recovers all the results
of the previous section at low fields, and also allows one to study
the problem well away from linear response.

We examine the electron-transport equation in a 1D uniform wire.
In steady state, with driving field $E$, the electron
distribution function $f_k$ in wave-vector
space $\{{k\}}$ obeys

$$
{eE\over \hbar}
{\partial f_k\over \partial k}
=
-{1\over \tau_{\rm in}(\varepsilon_k)}
{\left( f_k -
{{\langle \tau_{\rm in}^{-1} f \rangle}\over 
 {\langle \tau_{\rm in}^{-1} f^{\rm eq} \rangle}}
f^{\rm eq}_k
\right)}
-{1\over \tau_{\rm el}(\varepsilon_k)}
{ {f_k - f_{-k}}\over 2 }.
\eqno(12)
$$

\noindent
The scattering times $\tau_{\rm in}(\varepsilon_k)$
and $\tau_{\rm el}(\varepsilon_k)$
are in general energy-dependent.
Equation (12) has some essential properties:

\item{1.}
{\it Thermodynamic consistency}.
Only one chemical potential enters the problem,
through the equilibrium Fermi distribution

$$
f^{\rm eq}_k = 1/{\{ 1 +
\exp[(\varepsilon_k + \varepsilon_i - \mu)/k_{\rm B}T] \}}
$$

\item{}
(where $\varepsilon_i$ is the threshold energy
for the $i$th 1D subband and $k_{\rm B}T$ is the thermal energy).
This is the only place at which dependence on the chemical
potential $\mu$ comes in. 

\item{2.}
{\it Microscopic conservation}. The leading,
inelastic collision term
on the right of Eq. (12) has a restoring contribution
proportional to the expectation

$$
{\langle \tau_{\rm in}^{-1} f \rangle}
\equiv \int^{\infty}_{-\infty}
{2dk\over 2\pi} \tau_{\rm in}^{-1}(\varepsilon_k) f_k.
$$

\item{}
The inelastic collision term respects continuity
and  gauge invariance.
Last, the second term on the right of Eq. (12)
represents the elastic collisions, which work to restore
symmetry to the nonequilibrium distribution $f_k$.

The equation can be solved systematically. For the special
case that the collision
times are independent of the electronic band energy
$\varepsilon_k$, there is an exact solution
(not only for $f_k$ but also for the much
richer nonequilibrium current correlation, ${\cal C}_{vv}(t)$).
The response behavior parallels that for
the Kubo analysis. Indeed, all of the results Eqs. (5)-(11)
are recovered[19].

\vskip12 truept
\centerline{\it d. A Worked Example}
\vskip 8 truept

\topinsert
\vskip -0.30truecm
\input psfig.sty
\centerline{\hskip10mm\psfig{figure=cmt26_fig1.ps,height=10.0truecm}}
%
\vskip -2.00truecm

\noindent
{\bf Figure 1.}
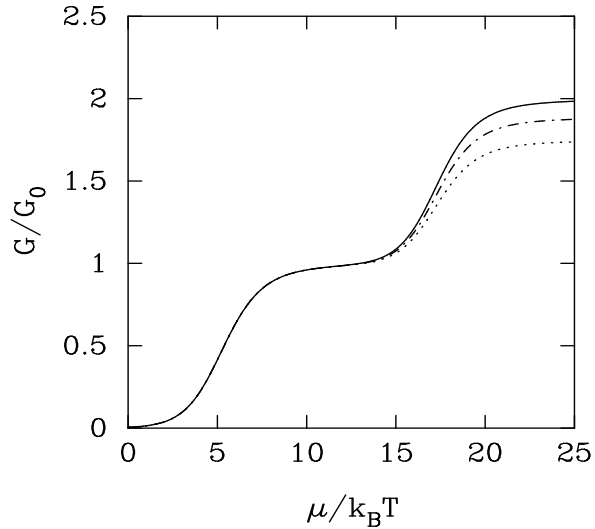
Conductance of a one-dimensional ballistic wire, computed from
Eq. (14) within the quantum kinetic model. We show $G$ as a function
of chemical potential $\mu$. The conductance is in units
of the Landauer quantum, and $\mu$ is in units of
thermal energy. $G$ exhibits strong shoulders as $\mu$ crosses
the two subband energy thresholds in succession. (These are set at
$\varepsilon_1 = 5k_{\rm B}T$ and $\varepsilon_2 = 17k_{\rm B}T$
in this simulation.) Well above each threshold, the electrons
in that subband are degenerate. The conductance tends to a
well defined quantized plateau.
Solid line: $G$ for an ideal ballistic channel: the Landauer limit.
Dot-dashed line: the collision-time ratio $\tau_{\rm in}/\tau_{\rm el}$
of the upper subband is set to the non-ideal value of 0.8.
Note how the increased inelastic scattering brings down the
upper plateau. Dotted line: as above, with
$\tau_{\rm in}/\tau_{\rm el} = 0.6$. The departure from ideality
is more pronounced. Such effects cannot be predicted from the
conventional phenomenology of the Landauer formula.
\vskip 8truept
\endinsert
\hsize = 15.5truecm

We end with a presentation of the conductance obtained from
our kinetic model. First, recall that the common derivation
of the Landauer conductance assumes a highly degenerate
electron band; that is, we are in the zero-temperature
limit. If the band is even marginally full, then the full
$G$ comes out. If the band is empty (the only other possibility
at zero temperature), then there is no conduction and $G = 0$.

In a real experimental situation at finite temperature, such as
in de Picciotto {\it et al}.[14] or van Wees {\it et al}.[15],
the carrier density in a 1D subband is controlled via
a gate above the wire. The electron population duly undergoes
a continuous transition, from a low-density classical regime to
a high-density Fermi-Dirac one. While there is no provision
for this classical-to-quantum crossover in the standard
treatments of the Landauer conductance, it is no problem
at all within the Kubo or kinetic treatment.

It is sufficient to note that, classically, the elastic
mean free path is a function, not of the Fermi velocity,
but of the thermal velocity $v_{\rm th} = \sqrt{ 2k_{\rm B}T/m^*}$.
In general, we must replace $\tau_{\rm el}$ with the expression

$$
\tau_{\rm el}(n, T) = {L\over {\overline v}}
\equiv L{n\over {\langle 2|v| f^{\rm eq} \rangle}}.
\eqno(13)
$$ 

\noindent
In the classical limit, ${\overline v} = v_{\rm th}$. In the
degenerate limit, ${\overline v} = v_{\rm F}$, as in Eq. (10).
We can then extend Eq. (8) for $G$ to all accessible regimes
of density at finite temperature:

$$
G = G_0 {\left( {hn\over 2m^*{\overline v}} \right)}
{\left( 1 - { 1\over {1 + \tau_{\rm in}/\tau_{\rm el}} } \right)},
\eqno(14)
$$

\noindent
where $G_0 = 2e^2/h$ is the Landauer conductance quantum,
and where $v_{\rm F}$ has been replaced with its equivalent
expression in 1D: $v_{\rm F} = \hbar k_{\rm F}/m^* = hn/4m^*$.

When the system is at low density ($\mu - \varepsilon_i \ll k_{\rm B}T$)
the conductance vanishes with $n$. When the system is degenerate
($\mu - \varepsilon_i \gg k_{\rm B}T$) the conductance reaches
a plateau, which is {\it ideally quantized}
when $\tau_{\rm in} =\tau_{\rm el}$. In between, it rises smoothly
as the chemical potential (and density) is swept from much below
the subband threshold $\varepsilon_i$ to much above it.
The result is depicted in Figure 1.

In Fig. 1 we see the conductance of a 1D ballistic wire
computed from Eq. (14), with full temperature dependence, as
a function of chemical potential. The threshold steps at
the two subband plateaux are clear. Also clear is the
progressive loss of ideality as the inelasticity in the
problem is increased. Nonetheless, the characteristic Landauer
steps survive robustly, even when the height of the steps no
longer corresponds to perfect ballistic transport inside the
body of the wire.

One principal conclusion stands out. In an open mesoscopic ballistic
conductor, the close interplay of elastic and dissipative
scattering dominates the form and behavior of the
conductance. On its own, neither collision mode can encapsulate
the relevant physics. {\it They must be allowed to act in concert,
as they do in nature}.

\vskip 20 truept
\centerline{\bf 5. EPILOG}
\vskip 12 truept
 
In this paper we have shown that the Landauer quantized-conductance
formula, foundational to so much of mesoscopic transport,
possesses a microscopic validity and scope well beyond the popular
rationale in which it has been clothed for so long. For an open
mesoscopic system, there is no escaping the
direct -- indeed vital -- interplay between elastic one-body
scattering on the one hand, and inelastic many-body scattering
on the other. Theories that favor the former at the expense
of the latter, without adducing valid microscopic reasons for
doing so, court a serious distortion of the physics.

Were it not for Landauer's intuition and his formula, mesoscopics in the
last two decades would have fared very differently. His legacy has truly
been one of unprecedented progress. It would therefore be a cause for
concern if, for the sake of fashion, inadequately reasoned casuistic
phenomenologies were to win out in the theoretical arena
over the microscopically based analytic methods first put in place by
Maxwell and Boltzmann, and culminating in the contributions of
Born, Fermi, Landau, and others.
Such a reversal of values does not serve progress in the long term.

One criterion matters for continued reliance on any mesoscopic
formula. It is not whether the formula is fashionable, nor whether
it is easily finessed for yet one more quick publication opportunity.
The {\it only} question that counts is whether the model has a
rational and clearly traceable origin within canonical
microscopics, the bedrock of modern condensed-matter physics.
If the answer is ``yes'', that formula will be far
more likely to keep on performing well.

\vskip 12 truept
\centerline{\bf REFERENCES}

\item{[1]}
R. Landauer, {\it IBM J. Res. Dev.} {\bf 1}, 223 (1957);
{\it Phil. Mag.}{\bf 21}, 863 (1970).

\item{[2]}
See e.g. A. D. Stone and A. Szafer,
{\it IBM J. Res. Dev.}{\bf 32}, 384 (1988).

\item{[3]}
R. Landauer, in {\it Coulomb and Interference
Effects in Small Electronic Structures},
ed. by D. C. Glattli, M. Sanquer and J. Tran Than Van
(Editions Fronti\`eres, Gif-sur-Yvette, 1994) p. 1.

\item{[4]} 
S. Doniach and E. H. Sondheimer,
{\it Green Functions for Solid State Physicists}
(W. A. Benjamin, Reading, MA, 1974).

\item {[5]}
F. Green and M. P. Das, in {\it Condensed Matter Theories Vol. 17},
ed. by M. P. Das and F. Green, (Nova Science Publ., New York, 2003).

\item {[6]}
M. Bonitz (editor), {\it Progress in Nonequilibrium Green's Functions}
(World Scientific, Singapore, 2000).

\item {[7]}
Y. Imry and R. Landauer, {\it Rev. Mod. Phys.} {\bf 71}, S306 (1999).

\item{[8]}
Y. Imry, {\it Introduction to Mesoscopic Physics} second edition
(Oxford University Press, Oxford, 2002).

\item {[9]}
D. K. Ferry and S. M. Goodnick,
{\it Transport in Nanostructures}
(Cambridge University Press, Cambridge, 1997).

\item{[10]}
S. Datta, {\it Electronic Transport in Mesoscopic Systems}
(Cambridge University Press, Cambridge, 1997).

\item{[11]}
J. H. Davies, {\it The Physics of Low Dimensional Semiconductors: an
Introduction}, (Cambridge University Press, Cambridge, 1998).

\item{[12]}
A. Kamenev and W. Kohn,
{\it Phys. Rev. B} {\bf 63}, 155304 (2001).

\item{[13]}
P. Dample, T. Rakshit, M. Paulsson, and S. Datta,
{\it Preprint} cond-mat/0206328.

\item{[14]}
R. de Picciotto, H. L. Stormer, L. N. Pfeiffer,
K. W. Baldwin, and K. W. West, {\it Nature} {\bf 411}, 51 (2001);
see also A. Chang, {\it Nature} {\bf 411}, 39 (2001).

\item{[15]}
B. J. van Wees, J. van Houten, C. W. J. Beenaker,
J. G. Williams, L. P. Kouwenhoven, D. van der Marel,
and C. T. Foxon, {\it Phys. Rev. Lett.} {\bf 60}, 848 (1988).

\item{[16]}
R. Kubo, M. Toda, and N. Hashitsume,
{\it Statistical Physics II: Nonequilibrium Statistical Mechanics},
second edition (Springer, Berlin, 1991).

\item{[17]}
F. Green and M. P. Das,
{\it J. Phys.: Condens. Matter} {\bf 12}, 5233 (2000).

\item{[18]}
F. Green and M. P. Das,
{\it J. Phys.: Condens. Matter} {\bf 12}, 5251 (2000).

\item{[19]}
F. Green and M. P. Das,
{\it Fluctuation and Noise Letters} {\bf 1}, C21 (2001)

\end

\end{document}